\documentclass[a4paper,11pt]{article}
\pdfoutput=1 
\usepackage{jcappub} 
\usepackage{gensymb}
\usepackage[]{algorithm2e}
\usepackage{subcaption}
\usepackage{sidecap}
\usepackage{aasmacros}

\usepackage{pgfplots}
\usepackage{booktabs}
\usepackage{cleveref}
\usepackage{bm}
\usepackage{placeins}
\usepackage{multirow}
\usepackage{lipsum}
\usepackage{lettrine}
\usepackage{mathtools}
\usepackage[colorinlistoftodos,prependcaption,textsize=tiny,textwidth=20mm]{todonotes}

\usepackage{lineno}

\usepackage[T1]{fontenc} 

\newcommand{\x}{{\bf x}}

\DeclareMathOperator{\X}{\textbf{X}}

\DeclareMathOperator{\z}{\textbf{z}}
\newcommand{\E}{\mathbb{E}}

\definecolor{airforceblue}{rgb}{0.36, 0.54, 0.66}
\definecolor{applegreen}{rgb}{0.55, 0.71, 0.0}
\definecolor{awesome}{rgb}{1.0, 0.13, 0.32}
\definecolor{bananayellow}{rgb}{1.0, 0.88, 0.21}

\notoc

\usepackage{array}
\newcolumntype{L}[1]{>{\raggedright\let\newline\\\arraybackslash\hspace{0pt}}m{#1}}
\newcolumntype{C}[1]{>{\centering\let\newline\\\arraybackslash\hspace{0pt}}m{#1}}
\newcolumntype{R}[1]{>{\raggedleft\let\newline\\\arraybackslash\hspace{0pt}}m{#1}}
\newcolumntype{P}[1]{>{\centering\arraybackslash}p{#1}}
\title{\boldmath Fast Cosmic Web Simulations with Generative Adversarial Networks}

 \author[a]{A.C. Rodr\'iguez,}
 \author[b]{T. Kacprzak,}
 \author[c]{A. Lucchi,}
 \author[b]{A. Amara,}
 \author[b]{R. Sgier,}
 \author[b]{J. Fluri,}
 \author[c]{T. Hofmann,}
 \author[b]{and A. R\'{e}fr\'{e}gier}
 \affiliation[a]{Photogrammetry \& Remote Sensing, Eidgen\"ossische Technische Hochschule Z\"urich,\\Stefano-Franscini-Platz 5, 8093 Z\"{u}rich, Switzerland}
 \affiliation[b]{Institute for Particle Physics and Astrophysics, Eidgen\"ossische Technische Hochschule Z\"urich,\\Wolfgang-Pauli-Str. 27, 8093 Z\"{u}rich, Switzerland}
 \affiliation[c]{Department of Computer Science, Eidgen\"ossische Technische Hochschule Z\"urich,\\Universit\"atstrasse 6, 8092 Z\"{u}rich, Switzerland}

\emailAdd{andres.rodriguez@geod.baug.ethz.ch}
\emailAdd{tomasz.kacprzak@phys.ethz.ch}
\emailAdd{aurelien.lucchi@inf.ethz.ch}
\emailAdd{adam.amara@phys.ethz.ch}
\emailAdd{rsgier@phys.ethz.ch}
\emailAdd{janis.fluri@phys.ethz.ch}
\emailAdd{thomas.hofmann@inf.ethz.ch}
\emailAdd{alexandre.refregier@phys.ethz.ch}


\abstract{
Dark matter in the universe evolves through gravity to form a complex network of halos, filaments, sheets and voids, that is known as the cosmic web.
Computational models of the underlying physical processes, such as classical N-body simulations, are extremely resource intensive, as they track the action of gravity in an expanding universe using billions of particles as tracers of the cosmic matter distribution.
Therefore, upcoming cosmology experiments will face a computational bottleneck that may limit the exploitation of their full scientific potential.
To address this challenge, we demonstrate the application of a machine learning technique called Generative Adversarial Networks (GAN) to learn models that can efficiently generate new, physically realistic realizations of the cosmic web.
Our training set is a small, representative sample of 2D image snapshots from N-body simulations of size 500 and 100 Mpc.
We show that the GAN-generated samples are qualitatively and quantitatively very similar to the originals.
For the larger boxes of size 500 Mpc, it is very difficult to distinguish them visually.
The agreement of the power spectrum $P_k$ is 1-2\% for most of the range, between $k=0.06$ and $k=0.4$.
For the remaining values of $k$, the agreement is within 15\%, with the error rate increasing for $k>0.8$.
For smaller boxes of size 100 Mpc, we find that the visual agreement to be good, but some differences are noticable.
The error on the power spectrum is of the order of 20\%.
We attribute this loss of performance to the fact that the matter distribution in 100 Mpc cutouts was very inhomogeneous between images, a situation in which the performance of GANs is known to deteriorate.
We find a good match for the correlation matrix of full $P_k$ range for 100 Mpc data and of small scales for 500 Mpc, with $\sim$20\% disagreement for large scales.
An important advantage of generating cosmic web realizations with a GAN is the considerable gains in terms of computation time. Each new sample generated by a GAN takes a fraction of a second, compared to the many hours needed by traditional N-body techniques.
We anticipate that the use of generative models such as GANs will therefore play an important role in providing extremely fast and precise simulations of cosmic web in the era of large cosmological surveys, such as Euclid and Large Synoptic Survey Telescope (LSST).

}
\graphicspath{{figures/}}
\begin{document}
\maketitle
\flushbottom
\newpage

\section{Introduction}

The large scale distribution of matter in the universe takes the form of a complicated network called the \emph{cosmic web} \citep{Bond1996how,Coles2000characterizing,Foreroromero2009dynamical,Dietrich2012filament,Liebskind2017tracing}.
The properties of this distribution contain important cosmological information used to study the nature of dark matter, dark energy, and the laws of gravity \citep{Des2017dark,Hildebrandt2017kids,Joudaki2017kids450}, as different cosmological models give rise to dark matter distributions with different properties.
Simulations of these cosmic structures \citep{Springel2005cosmological,Potter2016pkdgrav3} play a fundamental role in understanding cosmological measurements~\citep{Fosalba2015mice2,Busha2013catalog}.
These simulations are commonly computed using \emph{N-body} techniques, which represent the matter distribution as a set of particles that evolve throughout cosmic time according to the underlying cosmological model and the laws of gravity.
Creating a single N-body simulation requires the use of large computational resources for a long period of time such as days or weeks \citep{Teyssier2009fullsky,Boylankolchin2009resolving}.
Furthermore, reliable measurements of cosmological parameters typically require a large number of simulations of various cosmological models \citep{Harnoisderaps2015simulations,Kacprzak2016cosmology}.
This creates a strong need for fast, approximate approaches for generating simulations of cosmic web \citep{Heitmann2010coyote1,Heitmann2009coyote2,Lawrence2010coyote3,Lin2015new,Howlett2015lpicola}.

Here we demonstrate the possibility of using deep generative models to synthesize samples of the cosmic web.
Deep generative models~\citep{Kingma2013autoencoding, Goodfellow2014generative} are able to learn complex distributions from a given set of data, and then generate new, statistically consistent data samples.
Such a deep generative model can be trained on a set of N-body simulations.
Once the training is complete, the generative model can create new, random dark matter distributions that are uncorrelated to the training examples.
A practical advantage of using a generative model is that the generation process is extremely fast, thus giving us the ability to generate a virtually unlimited number of samples of the cosmic web.
Having access to such a large amount of simulations can potentially enable more reliable scientific studies and would therefore enhance our ability to understand the physics of the Universe.

In the last decade, deep learning approaches have achieved outstanding results in many fields, especially for computer vision tasks such as image segmentation or object detection~\citep{krizhevsky2012imagenet}.
Deep convolutional neural networks (DCNN) have also recently been used as data generating mechanisms. Here a latent random vector, typically a high-dimensional Gaussian, is passed through a DCNN in order to output images.
Generative Adversarial Networks (GAN) create such a model by adopting an adversarial game setting between two DCNN players, a generator and a discriminator.
The goal of the generator is to produce samples resembling the originals while the discriminator aims at distinguishing the originals from the fake samples produced by the generator.
The training process ends when a Nash equilibrium is reached, that is when no player can do better by unilaterally changing his strategy.

The rise of deep generative models has sparked a strong interest in the field of astronomy.
Deep generative models have been used to generate astronomical images of galaxies \citep{regier2015deep, ravanbakhsh2017enabling, Schawinski2017generative} or to recover certain features out of noisy astrophysical images \citep{Schawinski2017generative}.
GANs were recently applied to generating samples of projected 2D mass distribution, called convergence \citep{Mustafa2017creating}.
This approach can generate random samples of convergence maps, which are consistent with the original simulated maps according to several summary statistics.
The projection process, however, washes out the complex network structures present in the dark matter distribution.
Here, we instead focus on generating the structure of the cosmic web without projection, therefore preserving the ability of the generative model to create halos, filaments, and sheets.
We accomplish our goal by synthesizing thin slices of dark matter distribution which have been pixelised to create 2D images that serve as training data for a GAN model.

A demonstration of this method on 2D slices presents a case for the development of deep learning methods able to generate full, 3D dark matter distributions.
For cosmological applications, it may be more efficient to work with the full 3D matter distributions generated by a GAN, rather then 2D convergence maps.
For gravitational lensing, the convergence map depends on the input distribution of background galaxies \citep[see][for review]{Refregier2003weak}; the 3D matter distribution is projected onto the sky plane by integrating the mass in radial direction against a lensing kernel, which depends on distribution $n(z)$ of redshifts $z$ of background galaxies.
For most lensing studies, the uncertainty on $n(z)$ is large and is effectively marginalised over.
If the 3D distributions are simulated, then the projection can be done analytically, for a given $n(z)$ \citep{HarnoisDeraps2012gravitational,Sgier2018fastgeneration}.
For a 2D generative model, a separate GAN would have to be trained for each $n(z)$ distribution.
This may be particularly important for analyses beyond the power spectrum, such as peak statistics \citep{Dietrich2009cosmology,Kacprzak2016cosmology,Martinet2017kids450} or deep learning \citep{Schmelze2017cosmological,Gupta2018nongaussian}, which use simulations to predict both the signal and its uncertainty.
In this paper we demonstrate the feasibility of GAN-based methods for capturing the type of matter distributions characteristic for in N-body simulations.
As the development of 3D generative methods for N-body data is likely to be a very challenging due to scalability issues and memory requirements, we consider this to be an important step in asserting that this approach is worth pursuing further.

In learning the cosmic web structures, which are more feature-rich than projected convergence maps, we encountered and addressed several important challenges.
The first was to handle data with very large dynamic range of the data; the density in the images created from slices of N-body simulations span several orders of magnitude.
Secondly, we explored how {\em mode collapse}, a feature of GANs causing the model to focuses on a single local minimum, affects the quality of results \citep{Tolstikhin2017adagan,Metz2016unrolled,Salimans2016improved}.
As mode collapse is expected to depend on the degree of homogeneity between samples, we tested the performance of GANs for both large and small cosmological volumes, of size 500 and 100 Mpc;
the matter density distributions in large boxes are considerably more homogeneous than in small boxes.

Finally, expanding on the work of \citep{Mustafa2017creating}, we additionally evaluate the cross-correlations of the GAN-generated data with itself and the training set.
A high cross-correlation would be an indication of lack of independence between the generated samples, a feature which we would judge to be undesirable in this task.
\\

The paper is organised as follows.
In Section \ref{sec:gans} we describe the Generative Adversarial Networks.
Section \ref{sec:data} contains the information on N-body simulations used.
Our implementation of the algorithm is described in Section \ref{sec:implementation} and diagnostics used to evaluate its performance are detailed in Section \ref{sec:diagnostics}.
We present the results in Section \ref{sec:results} and conclude in Section \ref{sec:conclusions}.

\section{Generative Adversarial Networks}

\label{sec:gans}
The basic idea behind GANs consists in pairing-up two neural networks: a generator network $G$ and a discriminator network $D$.
These networks are trained in an adversarial game setting.
The discriminator $D: \x \mapsto [0; 1]$ tries to probabilistically classify a sample $\x$ as being real or fake.
On the other hand, the generator $G: \z \mapsto \x$ tries to generate samples that look like they were drawn from the true data distribution $p_\text{data}$.
This generator makes use of a random variable $\z$ drawn from a given prior $p_{\text{prior}}(\z)$ which is typically a Gaussian distribution.
Formally, the two networks $D$ and $G$ play the following two-player minimax game:

\begin{align}
\min_G \max_D [ V(D, G) ]  \\ \nonumber
V(D, G) \coloneqq \ &  \E_{\bm{x} \sim p_{\text{data}}(\bm{x})}[\log D(\bm{x})] \ + \E_{\bm{z} \sim p_{\text{prior}}(\bm{z})}[\log (1 - D(G(\bm{z})))],
\label{eq:minimaxgame-definition}
\end{align}
where $\E$ is the expectation function.
The standard GAN approach~\cite{Goodfellow2014generative} aims at finding a Nash Equilibrium of this objective by using gradient-based techniques in an alternating fashion, sometimes coupled with stabilization techniques~\cite{Gulrajani2017improved, roth2017stabilizing}. As shown in~\cite{Goodfellow2014generative}, for the Bayes-optimal discriminator $D(\bm{x})$, the objective in Eq.~\ref{eq:minimaxgame-definition} reduces to the Jensen-Shannon divergence between $p_{\text{data}}$ and the distribution induced by the generator. The work of~\cite{Nowozin2016fgan} later generalized this to a more general class of f-divergences. An alternative formulation proposed in~\cite{Arjovsky2017wgan} uses the Wasserstein-1 distance to measure how different the real and fake samples are. In this work we experimented with both the standard GAN approach as well as Wasserstein GAN. We found both approaches to produce similar results and here present the results for 500 Mpc using Wasserstein-1 distance and 100 Mpc using the standard GAN approach.

\section{N-body simulations data}
\label{sec:data}

We created N-body simulations of cosmic structures in boxes of size 100 Mpc and 500 Mpc with 512$^3$ and 1,024$^3$ particles respectively.
We used L-PICOLA \citep{Howlett2015lpicola} to create 10 independent simulation boxes for both box sizes.
The cosmological model used was $\Lambda$CDM (Cold Dark Matter) with Hubble constant $H_{0}=100$, $h=70 \ \mathrm{km} \ \mathrm{s}^{-1} \ \mathrm{Mpc}^{-1}$, dark energy density $\Omega_{\Lambda} = 0.72$ and matter density $\Omega_{m} = 0.28$.
We used the particle distribution at redshift $z=0$.
We cut the boxes into thin slices to create grayscale, two-dimensional images of the cosmic web.
This is accomplished by dividing the $x$-coordinates into uniform intervals to create 1,000 segments.

We then selected 500 non-consecutive slices and repeated this process for the $y$ and $z$ axes, which gave us 1,500 samples from each of the 10 realizations, yielding a total of $15,000$ samples as our training dataset.
We pixelised these slices into $256 \times 256$ pixel images.
The value at each pixel corresponded to its particle count.
After the pixelisation, the images are smoothed with a Gaussian kernel with standard deviation of one pixel.
This step is done to decrease the particle shot noise.

Most existing GAN architectures are designed for natural images and therefore require an RGB representation with 3 channels and integer values between 0 and 255.
We adapted the DCNN architecture to work on our grayscale, floating-point images.
We scaled the image values to lie in the interval $[-1, 1]$ as we empirically found this transformation to improve performance.
Once we have trained our GAN model, newly generated samples are transformed back to the original range using an inverse transformation.
The transformation between the original, smoothed image $x$ and the scaled image $s$ was chosen to be:
\begin{equation}
s(x) = \frac{2x}{(x+a)}  - 1
\label{eqn:transformation}
\end{equation}
where $a$ is a free parameter.
This transformation is non-linear, and similar in nature to a logarithm function.
This choice was motivated by the fact, that the cosmic web has a high dynamic range between empty regions of space (voids with no particles) and super-massive halos (with many, concentrated particles).
This non-linear transformation enhances the contrast on features of interest, namely the network structure of filaments, sheets and halos.
The parameter $a$ allows to control the median value of the images, and was fixed to $a=4$ throughout the experimental section.
Immediately after the generation of a new, synthetic image, we apply the inverse function $s^{-1}(x)$ to transform it to the original space.

In this paper we used L-PICOLA: a faster, but approximate simulator.
For a real application of our method a more precise simulator would be used, such as GADGET-2 \citep{Springel2005cosmological} or PkdGrav3 \citep{Potter2016pkdgrav3}.
Nevertheless, for the purpose of demonstration of performance of GANs, we consider L-PICOLA simulations to be sufficient.
We do not expect the results to differ much if GANs were trained on simulations generated using more precise codes.

\section{Implementation and training}
\label{sec:implementation}
We use a slightly modified version of the standard DCGAN architecture \cite{radford2015unsupervised}, which was shown to achieve good results on natural images, including various datasets such as LSUN-Bedrooms (3 million indoor bedrooms images) \cite{Fisher2015construction} or the celebrity face dataset (CelebA, 200000 28x28 pixel celebrity faces) \cite{Liu2015faceattributes}.

Table \ref{tab:dcganarchitecture} presents the details of the architecture used for our experiments.
We used similar architectures for both the discriminator and the generator, consisting of five convolutional layers.
The total number of trainable parameters in both networks is $3.2 \cdot 10^7$.
We trained the networks until we achieved convergence in terms of the discriminant score for the standard version and a stable distance between the generated and real images for Wasserstein-1.

A commonly faced problem when training GANs is a phenomenon called mode collapse \citep{Tolstikhin2017adagan,Metz2016unrolled,Salimans2016improved}, where the network focuses on a subset of the modes of the underlying data distribution. In these regions where the generator is fooling the discriminator well, the generator might converge to them, leaving out parts of regions of the target distribution. Wasserstein-1 loss, has some empirical evidence to prevent mode collapse but still suffers from it.

We addressed this problem by doing early stopping, effectively selecting the network parameters during the training process by choosing the network that displayed the best agreement in terms of the power spectrum statistics described in Section~\ref{sec:diagnostics}.
This happened after 17 and 21 epochs (one epoch consists of one full training cycle over the training set) for the 500 and 100 Mpc images respectively, which took 16.1 and 7 hours on a single GPU Nvidia GTX 1080 with 8GB.
Table \ref{tab:hyperparameters} presents the set of hyperparameters used in our results.

\section{Diagnostics}
\label{sec:diagnostics}

The diagnostic measures used in this work are: average histogram of pixel values in the images, average histogram of values of maxima (``peaks''), average auto power spectrum and the average cross-power spectrum of pairs of images within the sample.

Matter density distribution can be described as dimensionless over-density field in space $\delta(\boldsymbol x) = ( \rho(\boldsymbol x)-\bar \rho ) / \bar \rho$,
where $\rho(\boldsymbol x)$ is the matter density at position $\boldsymbol x$ and $\bar \rho$ is the mean density in the universe.
The cross power spectrum $P_{\delta_1\delta_2}$ of the matter over-densities is calculated as follows
\begin{equation}
\langle \tilde \delta_1(\boldsymbol\ell) \tilde \delta_2^{*}(\boldsymbol\ell) \rangle =
(2\pi)^2 \delta_D(\boldsymbol\ell-\boldsymbol\ell') P_{\times}(\ell).
\end{equation}

where $\tilde \delta_1(\boldsymbol \ell)$ and  $\tilde \delta_2(\boldsymbol \ell)$ are the Fourier transforms of two over-density maps at each logarithmically spaced Fourier bin $\boldmath\ell$,
and $\delta_D$ is the Dirac delta function.
To compute the auto power spectrum, we set $\tilde \delta_1(\boldsymbol \ell)=\tilde \delta_2(\boldsymbol \ell)$.

We compute both auto and cross power spectrum from 2D images using a discrete Fourier transform, followed by averaging over angles.

One of the popular alternatives to power spectrum for analysing matter density distribution is the \emph{peak statistics}.
These statistics capture non-Gaussian features present in the cosmic web and are commonly used on weak lensing data \citep{Martinet2017kids,Kacprzak2016cosmology}.
A ``peak'' is a pixel in the density map that is higher than all its immediate 24 neighbours. The peaks are then counted as a function of their height.

\section{Results}
\label{sec:results}

We focused our study on two simulation regimes: large-scale distribution, simulated in boxes of size 500 Mpc, and small-scale distribution, with boxes of size 100 Mpc.
For both configurations we ran 10 independent simulations.
From these boxes, we cut out a total of 15,000 thin, 2D slices for each box size.
We design a GAN model where both the discriminator and generator are deep convolutional neural networks.
These networks consists of 5 layers, with 4 convolutional layers using filter sizes of $5 \times 5$ pixels.

We trained the model parameters using {ADAM, a gradient based optimizer \citep{Diedrick2014adam}}, which yields a model that can generate new, random cosmic web images.
We assessed the performance of the generative model in several ways.
First, we performed a visual comparison of the original and synthetic images.
A quantitative assessment of the results was performed based on summary statistics commonly used in cosmology, described in Section \ref{sec:diagnostics}.
The angular power spectrum is a standard measure used for describing the matter distribution \citep{Kilbinger2015cosmology}.
Another important statistic used for cosmological measurements is the distribution of maxima in the density distribution, often called ``peak statistics'' \citep{Dietrich2009cosmology,Kacprzak2016cosmology}.
This statistic compares the number of maxima in the maps as a function of their values.
We also assessed the statistical independence of samples generated by GANs, as real cosmic structures are expected to be independent due to isotropy and homogeneity of the universe, unless they are physically close to each other.
To assess the independence of generated cosmic web distributions, we compare the cross-correlations between pairs of images.
Another statistic we used was the histogram of pixel values of N-body and GAN-generated images.
{Finally, we calculated the covariance between the power spectrum values at different $k$.}

\subsection{Large images of size 500 Mpc}

\begin{figure*}[h!]
	\centering
	\includegraphics[width=0.9\linewidth]{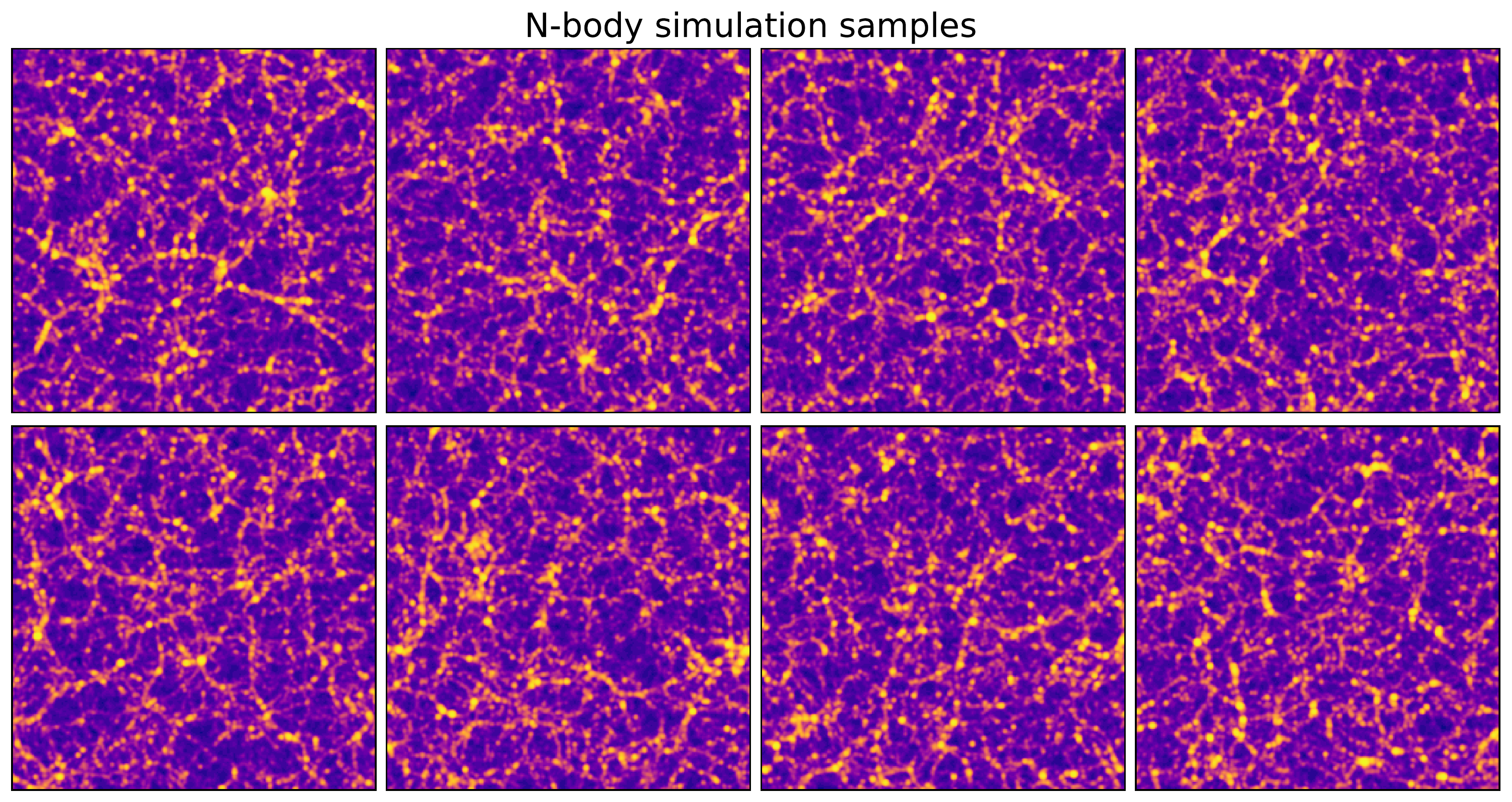}
	\includegraphics[width=0.9\linewidth]{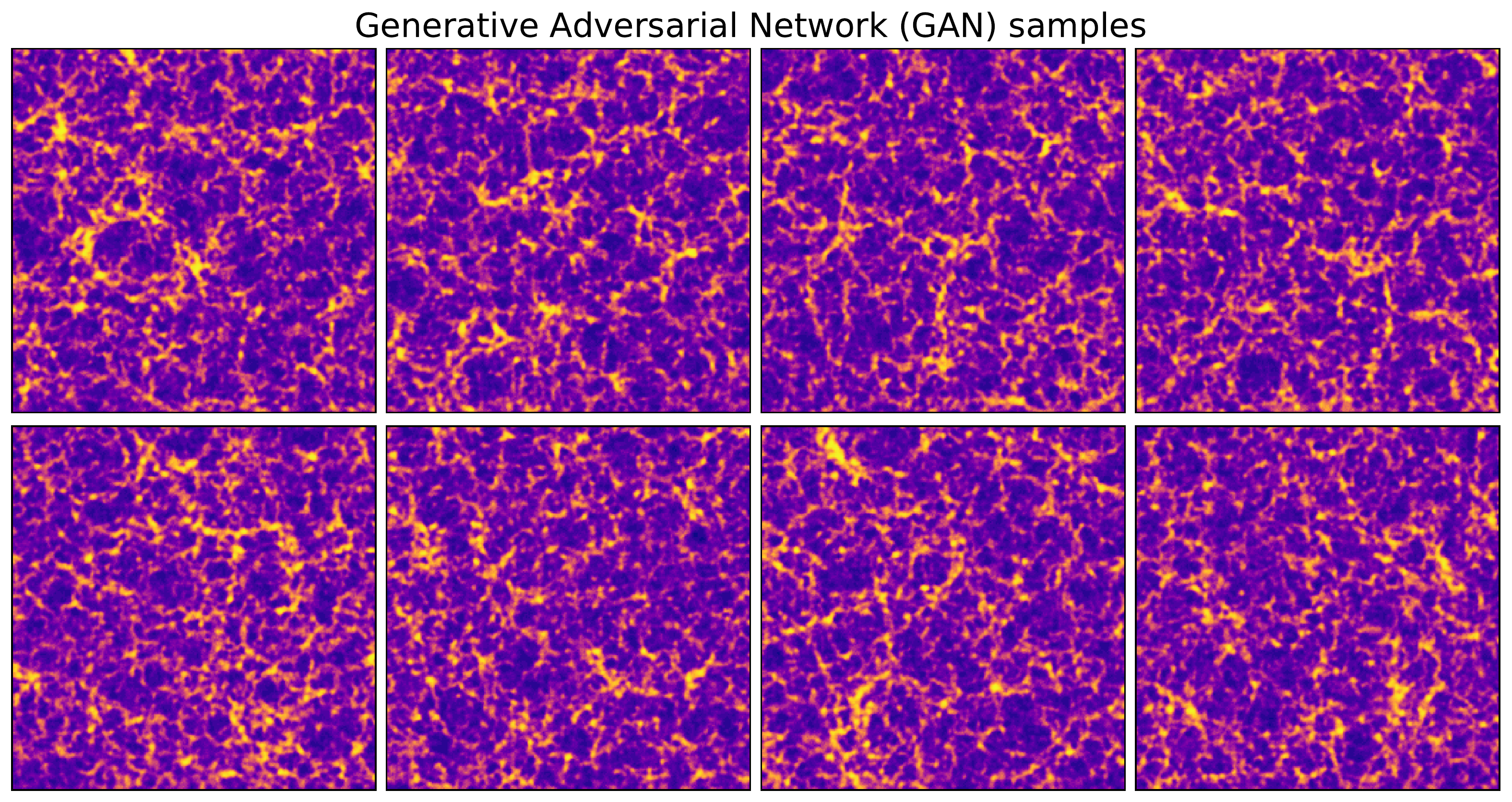}
	\caption{{Samples from N-body simulation and from GAN for the box size of 500 Mpc.}
		Note that the transformation in Equation~\ref{eqn:transformation} with $a=20$ was applied to the images shown above for better clarity.}
	\label{fig:gan-12-500}
\end{figure*}

Figure \ref{fig:gan-12-500} presents the original images (top) and synthesized images (bottom), for the 500 Mpc simulations.
The plotted images were transformed using Equation~\ref{eqn:transformation} to make it easier to assess the difference in the texture.
The cosmic web structure produced by the GAN model is visually very difficult to distinguish from the originals, even for human experts.
The GAN can capture the prominent features in this data, including halos and filaments.

\begin{figure*}[h!]
	\centering
	\includegraphics[width=0.8\linewidth]{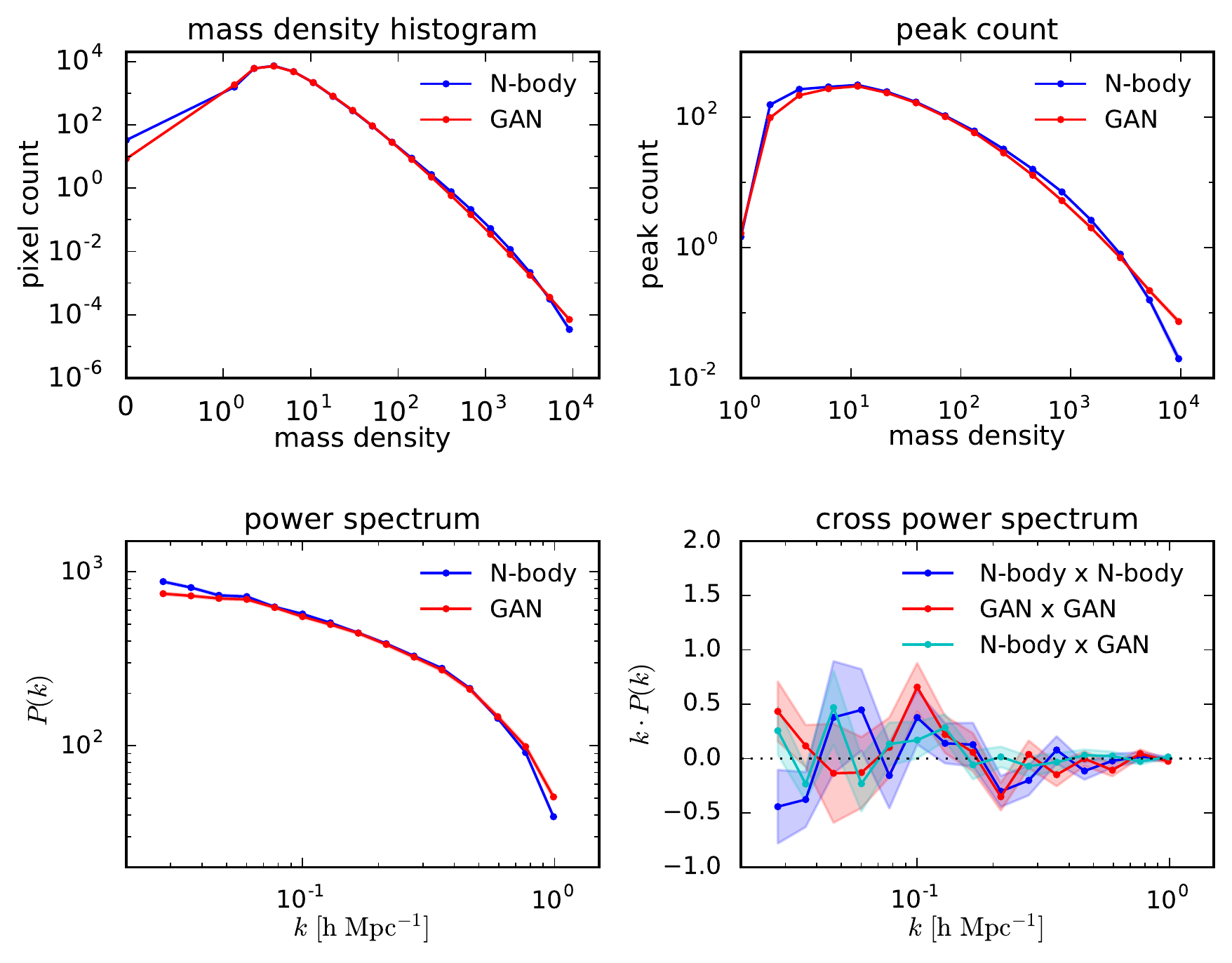}
	\caption{{Comparison of summary statistics between N-body and GAN simulations, for box size of 500 Mpc.}
		The statistics are: mass density histogram (upper left), peak count (upper right), power spectrum of 2D images (lower left) and cross power spectrum (lower right).
		The cross power spectrum is calculated between pairs N-body images (blue points), between pairs of GAN images (red points), and between pairs consisting of one GAN and one N-body image (cyan points).
		The power spectra are shown in units of $h^{-1} \ \rm{Mpc}$, where $h= \rm{H_0}/100$ corresponds to the Hubble parameter.
		The standard errors on the mean of the shown with a shaded region, and are too small to be seen for the first three panels.
	}
	\label{fig:metrics-112-500}
\end{figure*}

\begin{figure*}[h!]
	\centering
	\includegraphics[width=0.9\linewidth]{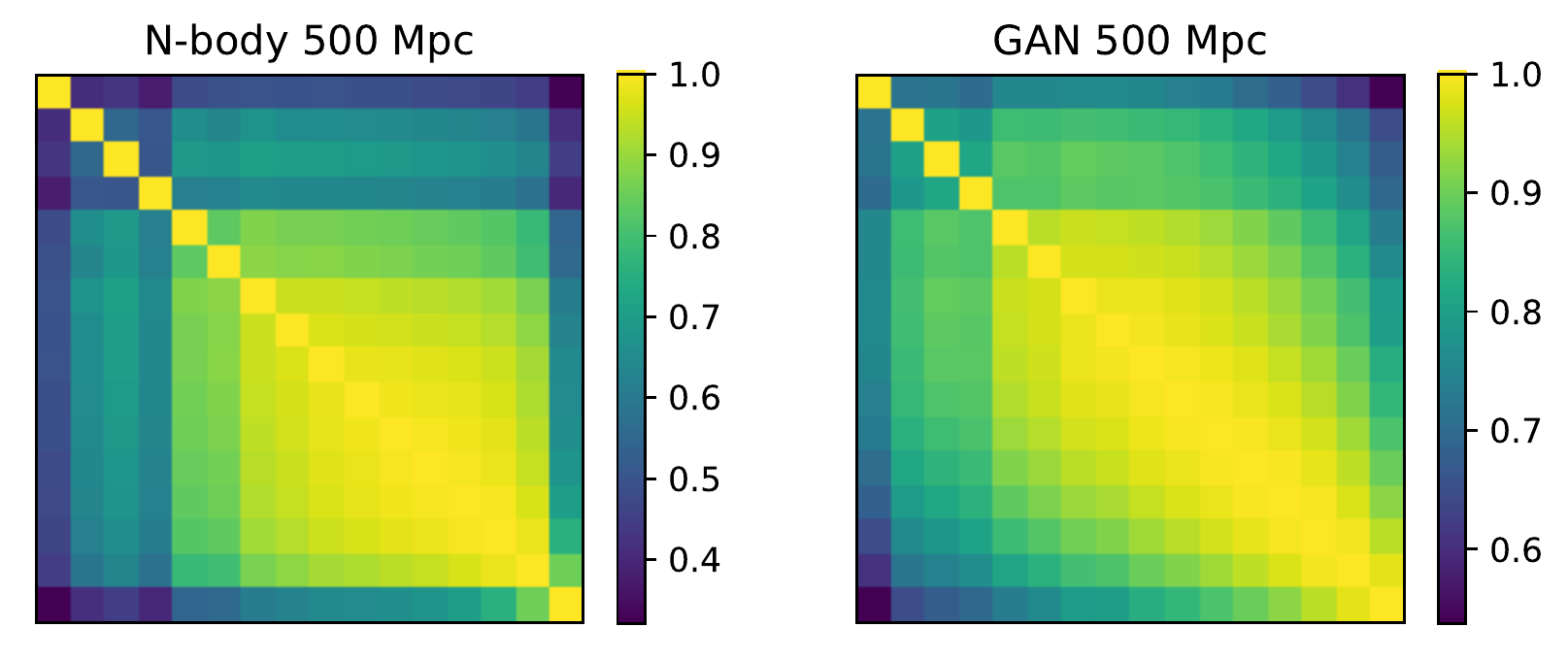}
	\includegraphics[width=0.9\linewidth]{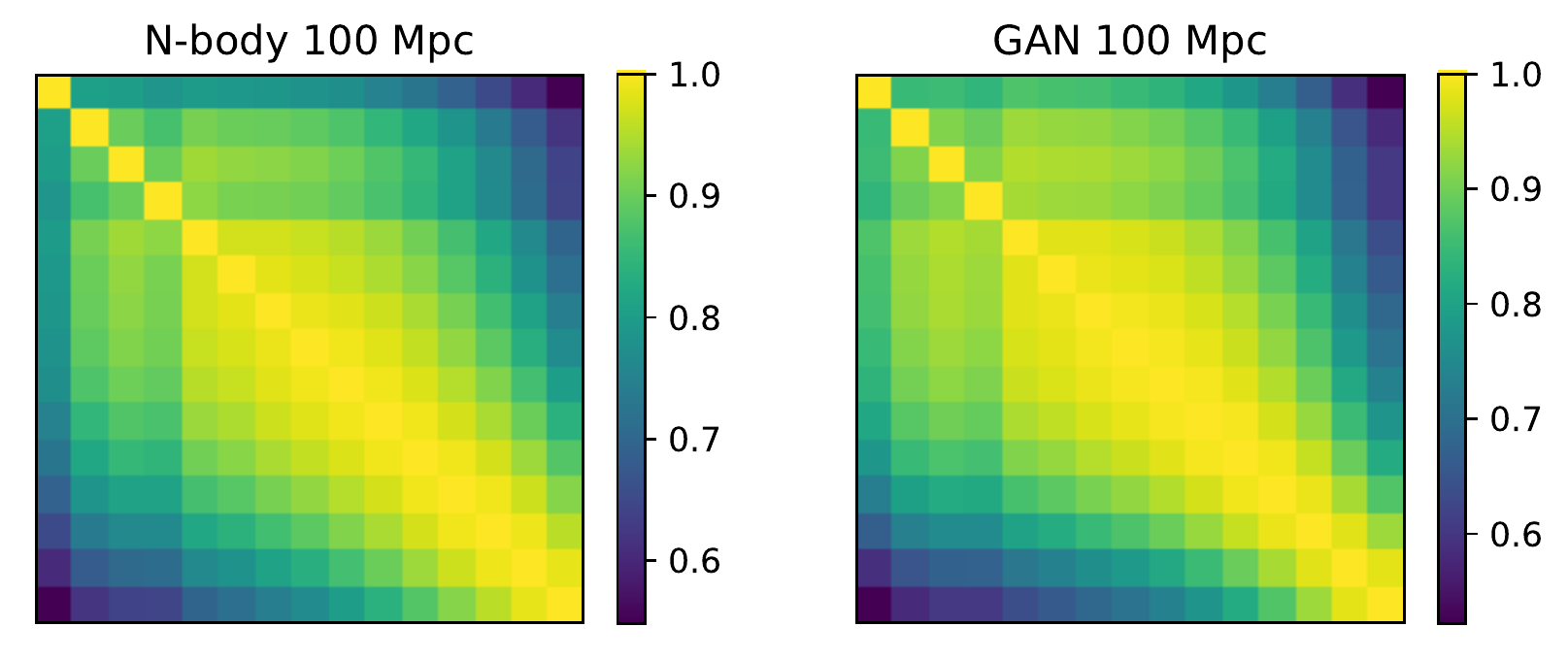}
	\caption{{Correlation matrix of $P(k)$ values from multiple images.}
		The correlation of the N-body and GAN-generated images is shown on the left and right panels, respectively.
		Top panels show the correlation for 500 Mpc images, and bottom panels for 100 Mpc.}
	\label{fig:cor}
\end{figure*}

Figure \ref{fig:metrics-112-500} shows the summary statistics for the original (blue lines) and GAN-synthesized (red lines) samples for 500 Mpc images.
Mass density histograms, shown in the top left panel, agree well throughout most of the range, except for very large densities.
Peak statistics, shown in top right panel, also agree well, although slightly worse than the density histograms, especially for higher mass ranges, where the error can reach $\sim$10\%.
The power spectrum is shown in the bottom left panel.
We focused on correlations at angular scales larger than a few Mpc,
as the current N-body simulations do not agree well in their predictions for smaller scales \citep{Schneider2016matter}.
We find that between $k$=0.06 and $k$=0.4 the agreement is 1-2\%, while the rest of the range agrees within 15\%, and for large $k>0.8$, the error starts to increase dramatically.
Finally, the bottom right panels show the average cross power spectra, with the coloured bands corresponding to the standard deviation calculated using all available image pairs.
As expected, the cross power spectrum of the original images is close to zero.
We do not find evident discrepancies in the cross power spectrum between pairs consisting of N-body- and GAN-generated image, as well as between pairs of GAN-generated images.
This indicates that the generated images can be also considered as uncorrelated realisations of cosmic web.
While the lack of cross correlation does not strictly imply independence, it assures that local structures are not consistently the same, and the data is not simply memorised and ``pasted'' in the same locations.
{Finally, the correlation matrix of power spectra at different values of $k$ is shown in the top panels of Figure~\ref{fig:cor}.
	For the 500 Mpc images, the structure of the correlation matrix for GAN is similar to N-body: more correlation is observed at small scales. The numerical agreement, however, is good only for small scales, with $\sim$5\% differences. For large scales, the errors reach 20\%.}

\subsection{Small images of size 100 Mpc}

\begin{figure*}[h!]
	\centering
	\includegraphics[width=0.9\linewidth]{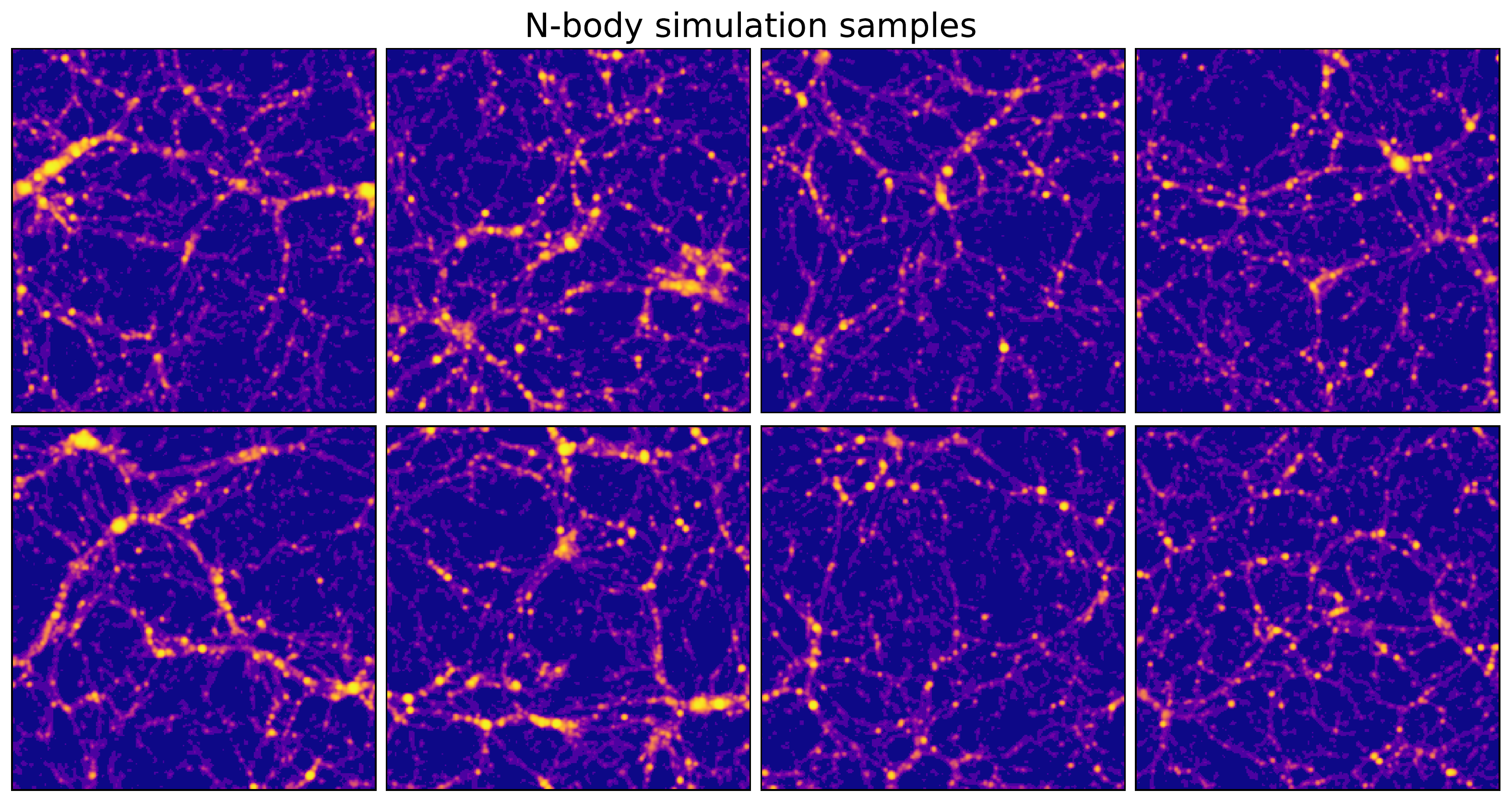}
	\includegraphics[width=0.9\linewidth]{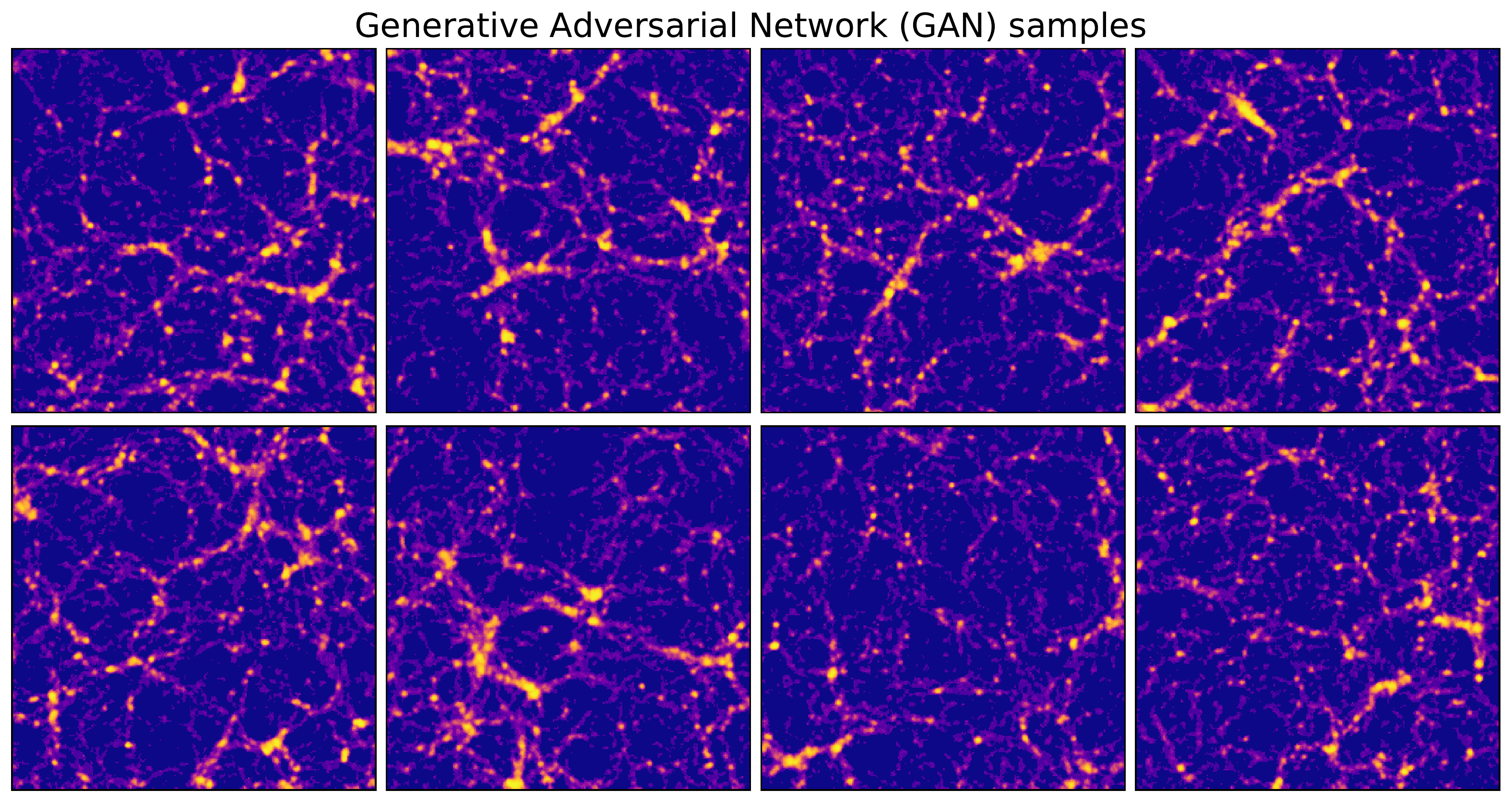}
	\caption{{Samples from N-body simulation and from GAN for the box size of 100 Mpc.}
		In this figure, transformation in Equation~\ref{eqn:transformation} with $a=7$ was applied.}
	\label{fig:gan-12-100}
\end{figure*}

The example density distributions from 100 Mpc data is shown in Figure \ref{fig:gan-12-100}.
These images are less homogeneous than the ones of size 500 Mpc.
The structures present in smaller images can vary from image to image: some may contain only empty space while some might be large structures.
The agreement between the real and GAN-generated images is still good, although it is possible to distinguish them visually.
Notably, the filaments do not look as distinct as in the real images.
Even thought the images are not homogeneous, the network can still capture this variability: it does generate images full of structures, as well as rather empty ones.
However, the proportions of these types of distributions among the generated samples may differ between real and GAN data.
These differences will manifest themselves in the quantitative comparison.

\begin{figure*}[h!]
	\centering
	\includegraphics[width=0.8\linewidth]{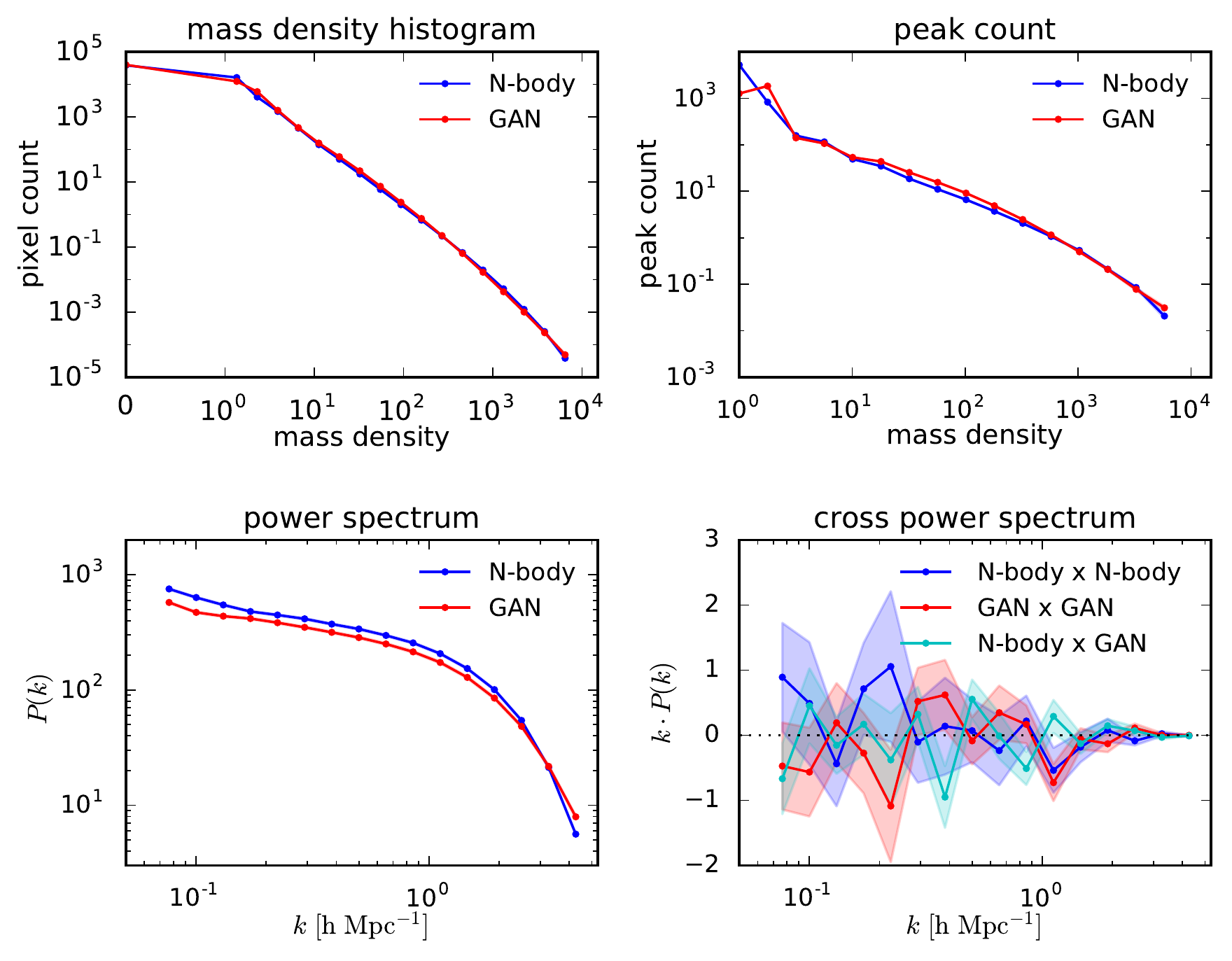}
	\caption{{Comparison of summary statistics between N-body and GAN simulations, for a box size of 100 Mpc.}
		The statistics are the same as in Figure~\ref{fig:metrics-112-500}.}
	\label{fig:metrics-112-100}
\end{figure*}

Figure \ref{fig:metrics-112-100} shows the summary statistics for 100 Mpc images.
The agreement between mass density histograms is good.
The difference in terms of peak statistics is on average small, although with deviations of $\sim$10 \% in the middle of the mass range.
The error on the power spectrum is much larger: between $k=0.13$ and $k=4$ there is a 20\% disagreement, and reaches 30\% outside that range.
From from $k>5$ the error becomes large.
Similarly to 500 Mpc images, we do not observe large discrepancies in the cross power spectrum between pairs of GAN generated images, as well as between GAN-real pairs.
{The agreement between the correlation matrix of the power spectra between N-body and GAN-generated is much better for 100 Mpc images.
	The differences are smaller than 5\% for most of the correlation matrix, as shown in the bottom panels in Figure~\ref{fig:cor}.}

\newpage
\section{Conclusion}
\label{sec:conclusions}
We demonstrated the ability of Generative Adversarial Models to learn the distribution describing the complex structures of the cosmic web.
We implemented a generative model based on deep convolutional neural networks, trained it on 2D images of cosmic web produced from N-body simulations, and used it to generate a synthetic cosmic web.
Our GAN-generated images are visually very similar to the ones from N-body simulations: the generative model managed to capture the complex structures of halos, filaments and voids.
We compared the GAN-generated images to the N-body originals using several summary statistics and found a good agreement.
Most notably, for 500 Mpc, the power agreement on power spectrum was very good: between $k=0.06$ and $k=0.4$ the level of 1-2\% is close to the requirements for precision cosmology \citep{Schneider2016matter}.
{The correlation matrices of $P(k)$ values had similar structures and agreed to around 5\% at small scales, but the GANs did not reproduce the large scale correlations well, with $\sim$20\% difference.}
While more work would be needed to improve this agreement further, this result is  promising for using GANs as emulators of mass density distributions for practical applications.

For 100 Mpc images the error on the power spectrum was larger, reaching 20\%.
We attribute this feature to the fact that images in the 100 Mpc sample are much more inhomogeneous than the 500 Mpc sample: some images contain dense regions with halos, and some relatively empty regions with few features.
We have seen empirically that this tends to induce a known phenomenon in GANs called mode collapse, where the training algorithm focuses on a {subset of the target distribution} .
This results in the model generating a few specific types of images, for example the ones with empty regions, more often than others.
We conclude that the application of GANs is suitable for large, homogeneous datasets.
For the type of inhomogeneous distributions appearing in the 100 Mpc sample, some techniques addressing mode-collapse~\cite{srivastava2017veegan, grnarova2017online} might be required if high-quality statistics are required.

An important advantage of the approach we presented here is that, once trained, it generates new samples in a fraction of a second on a modern Graphics Processing Unit (GPU).
Compared to a classical N-body technique, this constitutes a gain of several orders of magnitude in terms of simulation time.
The availability of this approach has the potential to dramatically reduce the computational burden required to acquire the data needed for most cosmological analyses.
Examples of such analyses include the computation of covariance matrices for cosmology with large scale structure \citep{Harnoisderaps2015simulations} or analyses using weak lensing shear peak statistics \citep{Dietrich2009cosmology}.
Generative methods may become even more important in the future; the need for fast N-body simulations is anticipated to grow in the era of large cosmological datasets obtained by the Euclid\footnote{\url{www.euclid-ec.org}} and LSST\footnote{\url{www.lsst.org}} projects.
The need for fast simulations will be amplified further by the emergence of new analysis methods, which can be based on advanced statistics \citep{Petri2013cosmology} or deep learning \citep{Schmelze2017cosmological}.
These methods aim to extract more information from cosmological data and often use large simulation datasets.
While we demonstrated the performance of GANs for 2D images using training on a single GPU, this approach can naturally be extended to generate 3D mass distributions \citep{ravanbakhsh2016estimating} for estimating cosmological parameters from dark matter simulations.

Finally, it would be interesting to explore how many simulations are needed to train a GAN model for a given precision requirement.
Another future direction would be to further explore the agreement between the original and GAN-generated images in terms of advanced statistics, such as for example 3-pt functions or Minkowski functionals.
Going beyond the cross-correlations to further tests for independence of the GAN-generated samples could also be of interest.
We leave this exploration to future work.

\acknowledgments
This work was support in part by grant number 200021\_169130 from the Swiss National Science Foundation.
We acknowledge the support of IT services of the Leonhard and Euler clusters at ETH Zurich.

\bibliography{References}{}

\begin{thebibliography}{10}

\bibitem{Bond1996how}
J.~R. {Bond}, L.~{Kofman}, and D.~{Pogosyan}.
\newblock {How filaments of galaxies are woven into the cosmic web}.
\newblock {\em \nat}, 380:603--606, April 1996.

\bibitem{Coles2000characterizing}
P.~{Coles} and L.-Y. {Chiang}.
\newblock {Characterizing the nonlinear growth of large-scale structure in the
  Universe}.
\newblock {\em \nat}, 406:376--378, July 2000.

\bibitem{Foreroromero2009dynamical}
J.~E. {Forero-Romero}, Y.~{Hoffman}, S.~{Gottl{\"o}ber}, A.~{Klypin}, and
  G.~{Yepes}.
\newblock {A dynamical classification of the cosmic web}.
\newblock {\em \mnras}, 396:1815--1824, July 2009.

\bibitem{Dietrich2012filament}
J.~P. {Dietrich}, N.~{Werner}, D.~{Clowe}, A.~{Finoguenov}, T.~{Kitching},
  L.~{Miller}, and A.~{Simionescu}.
\newblock {A filament of dark matter between two clusters of galaxies}.
\newblock {\em \nat}, 487:202--204, July 2012.

\bibitem{Liebskind2017tracing}
N.~I {Libeskind}, R.~{van de Weygaert}, M.~{Cautun}, B.~{Falck}, E.~{Tempel},
  T.~{Abel}, M.~{Alpaslan}, M.~A. {Aragoon-Calvo}, J.~E. {Forero-Romero},
  R.~{Gonzalez}, S.~{Gottloober}, O.~{Hahn}, W.~A. {Hellwing}, Y.~{Hoffman},
  B.~J.~T. {Jones}, F.~{Kitaura}, A.~{Knebe}, S.~{Manti}, M.~{Neyrinck}, S.~E.
  {Nuza}, N.~{Padilla}, E.~{Platen}, N.~{Ramachandra}, A.~{Robotham},
  E.~{Saar}, S.~{Shandarin}, M.~{Steinmetz}, R.~S. {Stoica}, T.~{Sousbie}, and
  G.~{Yepes}.
\newblock {Tracing the cosmic web}.
\newblock {\em ArXiv 1705.03021}, May 2017.

\bibitem{Des2017dark}
{DES Collaboration}.
\newblock {Dark Energy Survey Year 1 Results: Cosmological Constraints from
  Galaxy Clustering and Weak Lensing}.
\newblock {\em ArXiv 1708.01530}, August 2017.

\bibitem{Hildebrandt2017kids}
H~Hildebrandt, M~Viola, C~Heymans, S~Joudaki, K~Kuijken, C~Blake, T~Erben,
  B~Joachimi, D~Klaes, L~Miller, et~al.
\newblock Kids-450: Cosmological parameter constraints from tomographic weak
  gravitational lensing.
\newblock {\em Monthly Notices of the Royal Astronomical Society},
  465(2):1454--1498, 2017.

\bibitem{Joudaki2017kids450}
S.~{Joudaki}, A.~{Mead}, C.~{Blake}, A.~{Choi}, J.~{de Jong}, T.~{Erben},
  I.~{Fenech Conti}, R.~{Herbonnet}, C.~{Heymans}, H.~{Hildebrandt},
  H.~{Hoekstra}, B.~{Joachimi}, D.~{Klaes}, F.~{K{\"o}hlinger}, K.~{Kuijken},
  J.~{McFarland}, L.~{Miller}, P.~{Schneider}, and M.~{Viola}.
\newblock {KiDS-450: testing extensions to the standard cosmological model}.
\newblock {\em \mnras}, 471:1259--1279, October 2017.

\bibitem{Springel2005cosmological}
V.~{Springel}.
\newblock {The cosmological simulation code GADGET-2}.
\newblock {\em \mnras}, 364:1105--1134, December 2005.

\bibitem{Potter2016pkdgrav3}
Douglas Potter, Joachim Stadel, and Romain Teyssier.
\newblock Pkdgrav3: beyond trillion particle cosmological simulations for the
  next era of galaxy surveys.
\newblock {\em Computational Astrophysics and Cosmology}, 4(1):2, May 2017.

\bibitem{Fosalba2015mice2}
P.~{Fosalba}, E.~{Gazta{\~n}aga}, F.~J. {Castander}, and M.~{Crocce}.
\newblock {The MICE Grand Challenge light-cone simulation - III. Galaxy lensing
  mocks from all-sky lensing maps}.
\newblock {\em \mnras}, 447:1319--1332, February 2015.

\bibitem{Busha2013catalog}
M.~T. {Busha}, R.~H. {Wechsler}, M.~R. {Becker}, B.~{Erickson}, and A.~E.
  {Evrard}.
\newblock {Catalog Production for the DES Blind Cosmology Challenge}.
\newblock In {\em American Astronomical Society Meeting Abstracts \#221},
  volume 221, page 341.07, January 2013.

\bibitem{Teyssier2009fullsky}
R.~{Teyssier}, S.~{Pires}, S.~{Prunet}, D.~{Aubert}, C.~{Pichon}, A.~{Amara},
  K.~{Benabed}, S.~{Colombi}, A.~{Refregier}, and J.-L. {Starck}.
\newblock {Full-sky weak-lensing simulation with 70 billion particles}.
\newblock {\em \aap}, 497:335--341, April 2009.

\bibitem{Boylankolchin2009resolving}
M.~{Boylan-Kolchin}, V.~{Springel}, S.~D.~M. {White}, A.~{Jenkins}, and
  G.~{Lemson}.
\newblock {Resolving cosmic structure formation with the Millennium-II
  Simulation}.
\newblock {\em \mnras}, 398:1150--1164, September 2009.

\bibitem{Harnoisderaps2015simulations}
J.~{Harnois-D{\'e}raps} and L.~{van Waerbeke}.
\newblock {Simulations of weak gravitational lensing - II. Including finite
  support effects in cosmic shear covariance matrices}.
\newblock {\em \mnras}, 450:2857--2873, July 2015.

\bibitem{Kacprzak2016cosmology}
T~Kacprzak, D~Kirk, O~Friedrich, A~Amara, A~Refregier, L~Marian, JP~Dietrich,
  E~Suchyta, J~Aleksi{\'c}, D~Bacon, et~al.
\newblock Cosmology constraints from shear peak statistics in dark energy
  survey science verification data.
\newblock {\em Monthly Notices of the Royal Astronomical Society},
  463(4):3653--3673, 2016.

\bibitem{Heitmann2010coyote1}
K.~{Heitmann}, M.~{White}, C.~{Wagner}, S.~{Habib}, and D.~{Higdon}.
\newblock {The Coyote Universe. I. Precision Determination of the Nonlinear
  Matter Power Spectrum}.
\newblock {\em \apj}, 715:104--121, May 2010.

\bibitem{Heitmann2009coyote2}
K.~{Heitmann}, D.~{Higdon}, M.~{White}, S.~{Habib}, B.~J. {Williams},
  E.~{Lawrence}, and C.~{Wagner}.
\newblock {The Coyote Universe. II. Cosmological Models and Precision Emulation
  of the Nonlinear Matter Power Spectrum}.
\newblock {\em \apj}, 705:156--174, November 2009.

\bibitem{Lawrence2010coyote3}
E.~{Lawrence}, K.~{Heitmann}, M.~{White}, D.~{Higdon}, C.~{Wagner}, S.~{Habib},
  and B.~{Williams}.
\newblock {The Coyote Universe. III. Simulation Suite and Precision Emulator
  for the Nonlinear Matter Power Spectrum}.
\newblock {\em \apj}, 713:1322--1331, April 2010.

\bibitem{Lin2015new}
C.-A. {Lin} and M.~{Kilbinger}.
\newblock {A new model to predict weak-lensing peak counts. I. Comparison with
  N-body simulations}.
\newblock {\em \aap}, 576:A24, April 2015.

\bibitem{Howlett2015lpicola}
C.~{Howlett}, M.~{Manera}, and W.~J. {Percival}.
\newblock {L-PICOLA: A parallel code for fast dark matter simulation}.
\newblock {\em Astronomy and Computing}, 12:109--126, September 2015.

\bibitem{Kingma2013autoencoding}
D.~P {Kingma} and M.~{Welling}.
\newblock {Auto-Encoding Variational Bayes}.
\newblock {\em The International Conference on Learning Representations (ICLR),
  Banff, 2014, ArXiv 1312.6114}, December 2013.

\bibitem{Goodfellow2014generative}
I.~J. {Goodfellow}, J.~{Pouget-Abadie}, M.~{Mirza}, B.~{Xu}, D.~{Warde-Farley},
  S.~{Ozair}, A.~{Courville}, and Y.~{Bengio}.
\newblock {Generative Adversarial Networks}.
\newblock {\em ArXiv 1406.2661}, June 2014.

\bibitem{krizhevsky2012imagenet}
Alex Krizhevsky, Ilya Sutskever, and Geoffrey~E Hinton.
\newblock Imagenet classification with deep convolutional neural networks.
\newblock In {\em Advances in neural information processing systems}, pages
  1097--1105, 2012.

\bibitem{regier2015deep}
Jeffrey Regier, Jon McAuliffe, and Mr~Prabhat.
\newblock A deep generative model for astronomical images of galaxies.
\newblock In {\em NIPS Workshop: Advances in Approximate Bayesian Inference},
  2015.

\bibitem{ravanbakhsh2017enabling}
Siamak Ravanbakhsh, Francois Lanusse, Rachel Mandelbaum, Jeff~G Schneider, and
  Barnabas Poczos.
\newblock Enabling dark energy science with deep generative models of galaxy
  images.
\newblock In {\em AAAI}, pages 1488--1494, 2017.

\bibitem{Schawinski2017generative}
K.~{Schawinski}, C.~{Zhang}, H.~{Zhang}, L.~{Fowler}, and G.~K. {Santhanam}.
\newblock {Generative adversarial networks recover features in astrophysical
  images of galaxies beyond the deconvolution limit}.
\newblock {\em MNRAS}, 467:L110--L114, May 2017.

\bibitem{Mustafa2017creating}
M.~{Mustafa}, D.~{Bard}, W.~{Bhimji}, R.~{Al-Rfou}, and Z.~{Luki{\'c}}.
\newblock {Creating Virtual Universes Using Generative Adversarial Networks}.
\newblock {\em ArXiv 1706.02390}, June 2017.

\bibitem{Refregier2003weak}
A.~{Refregier}.
\newblock {Weak Gravitational Lensing by Large-Scale Structure}.
\newblock {\em \araa}, 41:645--668, 2003.

\bibitem{HarnoisDeraps2012gravitational}
J.~{Harnois-D{\'e}raps}, S.~{Vafaei}, and L.~{Van Waerbeke}.
\newblock {Gravitational lensing simulations - I. Covariance matrices and halo
  catalogues}.
\newblock {\em \mnras}, 426:1262--1279, October 2012.

\bibitem{Sgier2018fastgeneration}
R.~{Sgier}, A.~{R{\'e}fr{\'e}gier}, A.~{Amara}, and A.~{Nicola}.
\newblock {Fast Generation of Covariance Matrices for Weak Lensing}.
\newblock {\em ArXiv 1801.05745}, January 2018.

\bibitem{Dietrich2009cosmology}
J.~P. {Dietrich} and J.~{Hartlap}.
\newblock {Cosmology with the shear-peak statistics}.
\newblock {\em \mnras}, 402:1049--1058, February 2010.

\bibitem{Martinet2017kids450}
N.~{Martinet}, P.~{Schneider}, H.~{Hildebrandt}, H.~{Shan}, M.~{Asgari}, J.~P.
  {Dietrich}, J.~{Harnois-D{\'e}raps}, T.~{Erben}, A.~{Grado}, C.~{Heymans},
  H.~{Hoekstra}, D.~{Klaes}, K.~{Kuijken}, J.~{Merten}, and R.~{Nakajima}.
\newblock {KiDS-450: cosmological constraints from weak-lensing peak statistics
  - II: Inference from shear peaks using N-body simulations}.
\newblock {\em \mnras}, 474:712--730, February 2018.

\bibitem{Schmelze2017cosmological}
J.~{Schmelzle}, A.~{Lucchi}, T.~{Kacprzak}, A.~{Amara}, R.~{Sgier},
  A.~{R{\'e}fr{\'e}gier}, and T.~{Hofmann}.
\newblock {Cosmological model discrimination with Deep Learning}.
\newblock {\em ArXiv 1707.05167}, July 2017.

\bibitem{Gupta2018nongaussian}
A.~{Gupta}, J.~M. {Zorrilla Matilla}, D.~{Hsu}, and Z.~{Haiman}.
\newblock {Non-Gaussian information from weak lensing data via deep learning}.
\newblock {\em ArXiv 1802.01212}, February 2018.

\bibitem{Tolstikhin2017adagan}
I.~{Tolstikhin}, S.~{Gelly}, O.~{Bousquet}, C.-J. {Simon-Gabriel}, and
  B.~{Sch{\"o}lkopf}.
\newblock {AdaGAN: Boosting Generative Models}.
\newblock {\em ArXiv 1701.02386}, January 2017.

\bibitem{Metz2016unrolled}
L.~{Metz}, B.~{Poole}, D.~{Pfau}, and J.~{Sohl-Dickstein}.
\newblock {Unrolled Generative Adversarial Networks}.
\newblock {\em ArXiv 1611.02163}, November 2016.

\bibitem{Salimans2016improved}
Tim Salimans, Ian Goodfellow, Wojciech Zaremba, Vicki Cheung, Alec Radford,
  Xi~Chen, and Xi~Chen.
\newblock Improved techniques for training gans.
\newblock In {\em Advances in Neural Information Processing Systems 29}, pages
  2234--2242. Curran Associates, Inc., 2016.

\bibitem{Gulrajani2017improved}
I.~{Gulrajani}, F.~{Ahmed}, M.~{Arjovsky}, V.~{Dumoulin}, and A.~{Courville}.
\newblock {Improved Training of Wasserstein GANs}.
\newblock {\em ArXiv 1704.00028}, March 2017.

\bibitem{roth2017stabilizing}
Kevin Roth, Aurelien Lucchi, Sebastian Nowozin, and Thomas Hofmann.
\newblock Stabilizing training of generative adversarial networks through
  regularization.
\newblock In {\em Advances in Neural Information Processing Systems 30}, pages
  2018--2028. Curran Associates, Inc., 2017.

\bibitem{Nowozin2016fgan}
Sebastian Nowozin, Botond Cseke, and Ryota Tomioka.
\newblock f-gan: Training generative neural samplers using variational
  divergence minimization.
\newblock In {\em Advances in Neural Information Processing Systems 29}, pages
  271--279. Curran Associates, Inc., 2016.

\bibitem{Arjovsky2017wgan}
Martin Arjovsky, Soumith Chintala, and L{\'e}on Bottou.
\newblock {W}asserstein generative adversarial networks.
\newblock In Doina Precup and Yee~Whye Teh, editors, {\em Proceedings of the
  34th International Conference on Machine Learning}, volume~70 of {\em
  Proceedings of Machine Learning Research}, pages 214--223, International
  Convention Centre, Sydney, Australia, 06--11 Aug 2017. PMLR.

\bibitem{radford2015unsupervised}
Alec Radford, Luke Metz, and Soumith Chintala.
\newblock Unsupervised representation learning with deep convolutional
  generative adversarial networks.
\newblock {\em ArXiv 1511.06434}, 2015.

\bibitem{Fisher2015construction}
Fisher Yu, Yinda Zhang, Shuran Song, Ari Seff, and Jianxiong Xiao.
\newblock {LSUN:} construction of a large-scale image dataset using deep
  learning with humans in the loop.
\newblock {\em CoRR}, abs/1506.03365, 2015.

\bibitem{Liu2015faceattributes}
Ziwei Liu, Ping Luo, Xiaogang Wang, and Xiaoou Tang.
\newblock Deep learning face attributes in the wild.
\newblock In {\em Proceedings of the IEEE International Conference on Computer
  Vision}, pages 3730--3738, 2015.

\bibitem{Martinet2017kids}
N.~{Martinet}, P.~{Schneider}, H.~{Hildebrandt}, H.~{Shan}, M.~{Asgari}, J.~P.
  {Dietrich}, J.~{Harnois-D{\'e}raps}, T.~{Erben}, A.~{Grado}, C.~{Heymans},
  H.~{Hoekstra}, D.~{Klaes}, K.~{Kuijken}, J.~{Merten}, and R.~{Nakajima}.
\newblock {KiDS-450: Cosmological Constraints from Weak Lensing Peak Statistics
  - II: Inference from Shear Peaks using N-body Simulations}.
\newblock {\em ArXiv 1709.07678}, September 2017.

\bibitem{Diedrick2014adam}
Diederik~P. Kingma and Jimmy Ba.
\newblock Adam: {A} method for stochastic optimization.
\newblock {\em CoRR}, abs/1412.6980, 2014.

\bibitem{Kilbinger2015cosmology}
M.~{Kilbinger}.
\newblock {Cosmology with cosmic shear observations: a review}.
\newblock {\em Reports on Progress in Physics}, 78(8):086901, July 2015.

\bibitem{Schneider2016matter}
A.~{Schneider}, R.~{Teyssier}, D.~{Potter}, J.~{Stadel}, J.~{Onions}, D.~S.
  {Reed}, R.~E. {Smith}, V.~{Springel}, F.~R. {Pearce}, and R.~{Scoccimarro}.
\newblock {Matter power spectrum and the challenge of percent accuracy}.
\newblock {\em \jcap}, 4:047, April 2016.

\bibitem{srivastava2017veegan}
Akash Srivastava, Lazar Valkoz, Chris Russell, Michael~U Gutmann, and Charles
  Sutton.
\newblock Veegan: Reducing mode collapse in gans using implicit variational
  learning.
\newblock In {\em Advances in Neural Information Processing Systems}, pages
  3308--3318, 2017.

\bibitem{grnarova2017online}
Paulina Grnarova, Kfir~Y Levy, Aurelien Lucchi, Thomas Hofmann, and Andreas
  Krause.
\newblock An online learning approach to generative adversarial networks.
\newblock {\em arXiv preprint arXiv:1706.03269}, 2017.

\bibitem{Petri2013cosmology}
A.~{Petri}, Z.~{Haiman}, L.~{Hui}, M.~{May}, and J.~M. {Kratochvil}.
\newblock {Cosmology with Minkowski functionals and moments of the weak lensing
  convergence field}.
\newblock {\em \prd}, 88(12):123002, December 2013.

\bibitem{ravanbakhsh2016estimating}
Siamak Ravanbakhsh, Junier Oliva, Sebastian Fromenteau, Layne Price, Shirley
  Ho, Jeff Schneider, and Barnab{\'a}s P{\'o}czos.
\newblock Estimating cosmological parameters from the dark matter distribution.
\newblock In {\em International Conference on Machine Learning}, pages
  2407--2416, 2016.

\end{thebibliography}
\bibliographystyle{unsrt}

\appendix

\section{Network architecture and parameters}
\label{sec:appendix}
The architecture used in this work is shown in Table \ref{tab:dcganarchitecture}. See Section \ref{sec:implementation} for more details.
Table \ref{tab:hyperparameters} contains the list of hyperparameters used.

\begin{table*}[h!]
	\centering
	\begin{tabular}{cccc}
		\toprule
		Layer & Operation & Output & dimension \\
		\toprule
		\multicolumn{4}{l}{\textit{Discriminator}} \\
		\cmidrule{1-3}
		$\X$  &  & &  $ m \times 256 \times 256  \times 1 $ \\
		$h_0$  & conv  & LeakyRelu - BatchNorm &  $ m\times 128 \times 128  \times 64 $ \\
		$h_1$  & conv  & LeakyRelu - BatchNorm &  $ m \times 64 \times 64  \times 128 $ \\
		$h_2$  & conv  & LeakyRelu - BatchNorm &  $ m\times 32 \times 32  \times 256 $ \\
		$h_3$  & conv  & LeakyRelu - BatchNorm &  $ m \times 16 \times 16  \times 512 $ \\
		$h_4$  & linear  & sigmoid (identity) &  $ m \times 1 $ \\
		\multicolumn{4}{l}{\textit{Generator}} \\
		\cmidrule{1-3}
		$\z$  &  & &  $ m \times 200$ $(m \times 100)  $ \\
		$h_0$  & linear  & Relu - BatchNorm  & $ m \times 16 \times 16  \times 512 $ \\
		$h_1$  & deconv  & Relu - BatchNorm &  $ m \times 32 \times 32  \times 256 $ \\
		$h_2$  & deconv  & Relu - BatchNorm &  $ m \times 64 \times 64  \times 128 $ \\
		$h_3$  & deconv  & Relu - BatchNorm &  $ m \times 128 \times 128  \times 64 $ \\
		$h_4$  & deconv  & tanh &  $ m \times 256 \times 256  \times 1 $ \\
		\multicolumn{4}{l}{\scriptsize }
	\end{tabular}
	\caption{Architecture used in the discriminator and generator networks.
		We used a batch size of $m = 16$ samples. The neural network has $\sim$32 million trainable parameters.
		Parameters for our Wasserstein-1 distance implementation are shown in brackets.}
	\label{tab:dcganarchitecture}
\end{table*}

\begin{table*}[h!]
	\setlength{\extrarowheight}{0.1cm}
	\centering
	\begin{tabular}{p{2.9cm}P{2.3cm}P{2.3cm}p{7.5cm}}
		\toprule
		\multirow{2}{*}{Hyperparameter} & \multicolumn{2}{c}{GAN} & \multirow{2}{*}{Description}\\
		& Standard  & Wasserstein-1  & \\\toprule
		Batch size &  16 & 16  & Number of training samples used to compute the gradient at each update\\
		$\bm{z}$ dimension &  200 & 100 & Dimension of the gaussian prior distribution\\
		Learning rate D& $1 \cdot 10^{-5}$  & $1 \cdot 10^{-5}$  & {Discriminator} learning rate used by the \textsc{Adam} optimizer\\
		$\beta_1$ & 0.5 & 0.5 & Exponential decay for the \textsc{Adam} optimizer\\
		$\beta_2$ & 0.999 & 0.999 & Exponential decay for the \textsc{Adam} optimizer\\
		 {Learning rate G} & $1 \cdot 10^{-5}$ &$1 \cdot 10^{-8}$  & {Generator learning rate used by} the \textsc{Adam} optimizer\\
		Gradient penalty & - & 1,000 & Gradient penalty applied for Wasserstein-1 \\
		$a$ & 4 & 4 & Parameter in $s(x)$ to obtain the scaled images\\
	\end{tabular}
	\caption{Hyper-parameters used in our GAN implementations.
		\textsc{Adam} \cite{Diedrick2014adam} is the algorithm used to estimate the gradient in our models.}
	\label{tab:hyperparameters}
\end{table*}

\end{document}